\def\isarxiv{1} 

\ifdefined\isarxiv
\documentclass[11pt]{article}

\usepackage[numbers]{natbib}

\else
\documentclass{article}
\usepackage{microtype}
\usepackage{graphicx}
\usepackage{subfig}
\usepackage{hyperref}
\usepackage{icml2024}
\fi

\usepackage{amsmath}
\usepackage{amsthm}
\usepackage{amssymb}
\usepackage{algorithm}

\usepackage{algpseudocode}

\usepackage{grffile}
\usepackage{wrapfig,epsfig}
\usepackage{url}
\usepackage{xcolor}
\usepackage{epstopdf}

\usepackage{bbm}
\usepackage{dsfont}

\allowdisplaybreaks

\ifdefined\isarxiv

\let\C\relax
\usepackage{tikz}
\usepackage{hyperref}  
\hypersetup{colorlinks=true,citecolor=blue,linkcolor=blue} 
\usetikzlibrary{arrows}
\usepackage[margin=1in]{geometry}

\else

\fi

\theoremstyle{plain}
\newtheorem{theorem}{Theorem}[section]
\newtheorem{lemma}[theorem]{Lemma}

\newtheorem{fact}[theorem]{Fact}

\newtheorem{claim}[theorem]{Claim}

\theoremstyle{definition}
\newtheorem{definition}[theorem]{Definition}

\newcommand{\wt}{\widetilde}

\newcommand{\R}{\mathbb{R}}

\renewcommand{\tilde}{\wt}

\DeclareMathOperator*{\E}{{\mathbb{E}}}

\DeclareMathOperator*{\C}{\mathbb{C}}

\makeatletter
\newcommand*{\RN}[1]{\expandafter\@slowromancap\romannumeral #1@}
\makeatother

\usepackage{lineno}

\ifdefined\isarxiv

\else

\usepackage[textsize=tiny]{todonotes}

\icmltitlerunning{Quantum Speedup for Spectral Approximation of Kronecker Products}

\fi

\begin{document}

\ifdefined\isarxiv

\date{}

\title{Quantum Speedup for Spectral Approximation of Kronecker Products}
\author{
Yeqi Gao\thanks{\texttt{a916755226@gmail.com}. The University of Washington.}
\and
Zhao Song\thanks{\texttt{zsong@adobe.com}. Adobe Research.}
\and 
Ruizhe Zhang\thanks{\texttt{ruizhe@utexas.edu}. Simons Institute for the Theory of Computing.}.
}

\else

\twocolumn[
\icmltitle{Quantum Speedup for Spectral Approximation of Kronecker Products}



\icmlsetsymbol{equal}{*}

\begin{icmlauthorlist}
\icmlauthor{Firstname1 Lastname1}{equal,yyy}
\icmlauthor{Firstname2 Lastname2}{equal,yyy,comp}
\icmlauthor{Firstname3 Lastname3}{comp}
\icmlauthor{Firstname4 Lastname4}{sch}
\icmlauthor{Firstname5 Lastname5}{yyy}
\icmlauthor{Firstname6 Lastname6}{sch,yyy,comp}
\icmlauthor{Firstname7 Lastname7}{comp}
\icmlauthor{Firstname8 Lastname8}{sch}
\icmlauthor{Firstname8 Lastname8}{yyy,comp}
\end{icmlauthorlist}

\icmlaffiliation{yyy}{Department of XXX, University of YYY, Location, Country}
\icmlaffiliation{comp}{Company Name, Location, Country}
\icmlaffiliation{sch}{School of ZZZ, Institute of WWW, Location, Country}

\icmlcorrespondingauthor{Firstname1 Lastname1}{first1.last1@xxx.edu}
\icmlcorrespondingauthor{Firstname2 Lastname2}{first2.last2@www.uk}

\icmlkeywords{Machine Learning, ICML}

\vskip 0.3in
]



\printAffiliationsAndNotice{\icmlEqualContribution} 

\fi

\ifdefined\isarxiv
\begin{titlepage}
  \maketitle
  \begin{abstract}

Given its widespread application in machine learning and optimization, the Kronecker product emerges as a pivotal linear algebra operator. However, its computational demands render it an expensive operation, leading to heightened costs in spectral approximation of it through traditional computation algorithms.
Existing classical methods for spectral approximation exhibit a linear dependency on the matrix dimension denoted by $n$, considering matrices of size $A_1 \in \R^{n \times d}$ and $A_2 \in \R^{n \times d}$.
Our work introduces an innovative approach to efficiently address the spectral approximation of the Kronecker product $A_1 \otimes A_2$ using quantum methods. By treating matrices as quantum states, our proposed method significantly reduces the time complexity of spectral approximation to $O_{d,\epsilon}(\sqrt{n})$.

  \end{abstract}
  \thispagestyle{empty}
\end{titlepage}

{
}
\newpage

\else

\begin{abstract}

Given its widespread application in machine learning and optimization, the Kronecker product emerges as a pivotal linear algebra operator. However, its computational demands render it an expensive operation, leading to heightened costs in spectral approximation of it through traditional computation algorithms.
Existing classical methods for spectral approximation exhibit a linear dependency on the matrix dimension denoted by $n$, considering matrices of size $A_1 \in \R^{n \times d}$ and $A_2 \in \R^{n \times d}$.
Our work introduces an innovative approach to efficiently address the spectral approximation of the Kronecker product $A_1 \otimes A_2$ using quantum methods. By treating matrices as quantum states, our proposed method significantly reduces the time complexity of spectral approximation to $O_{d,\epsilon}(\sqrt{n})$.
\end{abstract}

\fi

\section{Introduction}\label{sec:intro}
Kronecker product is an important linear algebra operator. For two matrices $A\in \R^{n\times d}$ and $B\in \R^{n'\times d'}$, the Kronecker product $A\otimes B$ is an $nn'$-by-$dd'$ matrix such that for any $(i_1,j_1)\in [n]\times [n']$ and $(i_2,j_2)\in [d]\times [d']$,
\begin{align*}
    (A\otimes B)_{(i_1,j_1),(i_2,j_2)}:=A_{i_1,i_2}\cdot B_{j_1,j_2}.
\end{align*}
Due to its strong connection to tensors, it has been widely used in optimization and machine learning \cite{kb09,sdl+17,swz19}. 
However, the computational cost for the Kronecker product is very expensive. Naively, it takes $O(n^2d^2)$ time to compute the Kronecker product between two $n$-by-$d$ matrices. Thus, it has become an essential problem in sketching and sublinear-time algorithms to design efficient algorithms for the Kronecker product and related applications such as tensor regression, tensor low-rank approximation \cite{dssw18,swz19,djs+19}, etc. 

In this work, we study a fundamental problem of computing the spectral approximation for the Kronecker product. More specifically, given two matrices $A_1,A_2\in \R^{n\times d}$, let $A:=A_1\otimes A_2\in \R^{n^2\times d^2}$. The goal is to compute a matrix $B\in \R^{m\times d^2}$ with $m\ll n^2$ such that $B^\top B\approx_\epsilon A^\top A$. This can be regarded as a dimension reduction in that we reduce the number of rows in $A$ from $n^2$ to $m$. A direct application of this problem is to solve the tensor regression faster since it suffices to solve the ``sketched'' regression for the smaller-sized matrix $B$, and the spectral approximation guarantees that the solution will be close to the optimal solution of the original regression problem. Initiated by the seminal work of Pagh \cite{p13} the TensorSketch, several variations of tensor generalization of classical sketches have been proposed such as TensorSRHT \cite{akk+20,swyz21} (tensor generalization of SRHT \cite{ldfu13}), TensorSparse \cite{sxz22} (tensor generalization of \cite{nn13}). 
Unfortunately, the computational costs of these methods have at least a linear dependence on $n$, the number of rows. It is undesirable when the matrix is tall (also called over-constrained setting), i.e., $n\gg d$, which is indeed the case in a large number of practical applications, including linear regression \cite{cw17,nn13}, linear programming~\cite{blss20} neural network training \cite{szz24}. 
Therefore, it is natural to ask the following question:

\begin{center}
    {\it Can we compute the spectral approximation of $A_1\otimes A_2$ in $O_{d,\epsilon}(\sqrt{n})$-time\footnote{We use $O_g(f(n))$ to denote $f(n)\cdot g^{O(1)}$.}?}
\end{center}

\paragraph{Contributions.} In this work, we provide an affirmative answer to this question by proposing an efficient quantum algorithm to compute the spectral approximation of Kronecker products. Over the past few years, quantum computing has made remarkable progress: real devices with increasing capabilities have been built in the labs, and ``supremacy experiments'' have demonstrated quantum advantages over classical computations in some problems \cite{aru19,zwd20}. In particular, there has been a long line of work on quantum algorithms for optimization and machine-learning problems \cite{vgg17,wan17,kll19,am20,cll22,lz22,cmp23,dlt23}. One popular approach is quantum linear algebra (QLA), which treats matrices as quantum unitaries and vectors as quantum states \cite{p14,cgj18,gsl19,mrt21,tt24}. It can achieve exponential quantum speedups in some linear algebra tasks, such as solving linear systems, matrix-vector multiplication, etc. However, this approach may not be applied to sketching and spectral approximation due to the following limitations:
\begin{itemize}
    \item We expect classical outputs (vectors/matrices) for sketching in practical applications, while the QLA approach only encodes the outputs in quantum states. We need to apply quantum state tomography techniques \cite{cra10,acg23} to extract classical information from the quantum outputs, which corrupts the exponential quantum advantage.
    \item The costs of most QLA-based algorithms depend on several data-dependent parameters, including the condition number and Frobenius norm of the input matrices. However, the corresponding classical state-of-the-art algorithms do not depend on them. Thus, it is difficult to have a fair comparison between quantum and classical costs.  
\end{itemize}
This work is inspired by the recent result by \citet{ag23}, who developed a quantum matrix sketching method and used it to speed up linear programming (LP) solving. We generalize this method to sketch Kronecker products, which broadens its applications. Moreover, the output of our algorithm is purely classical, and the costs do not depend on any data-dependent parameter. Thus, in the tall matrix regime ($n\gg d$), our algorithm achieves a polynomial quantum speedup over the currently best classical algorithm. 
\subsection{Our Result}

We state our main result as follows
\begin{theorem}[Main Result]\label{lem:otimes_quantum}
Consider query access to matrix $A \in \R^{n^2 \times d^2}$ (where $A = A_1 \otimes A_2$, $A_1, A_2 \in \R^{n \times d}$ with row sparsity $r$). For any $\epsilon \in (0,1)$, there is a quantum algorithm that returns a diagonal matrix $D \in \R^{n^2 \times n^2}$ such that
\begin{itemize}
\item $\| D \|_0 = O(\epsilon^{-2} d^2 \log d)$
\item $  (1-\epsilon) A^\top A \preceq A^\top D^\top D A \preceq (1+\epsilon) A^\top A $
\item Each row in $DA$ is a scaling of one of rows in $A$.
\item $D \sim \mathsf{LS}(A)$.
\item It makes $\wt{O}(\sqrt{nd} / \epsilon )$ row queries to $A$
\item It takes $\wt{O}( r \sqrt{nd} / \epsilon + d^{\omega} )$ time
\item The success probability $0.999$
\end{itemize}
\end{theorem}
{\bf Roadmap.}
In Section~\ref{sec:related_work}, we present relevant literature pertaining to our contribution, encompassing leverage score and quantum sketching algorithms. We then delve into technical tools related to both classical mathematical and quantum properties in sampling, which will be instrumental in the analysis of spectral approximation. The Section~\ref{sec:gen_leverage_score} introduces the generalized leverage score, a key element facilitating the computation in our algorithm. The conclusive quantum algorithmic results are detailed in Section~\ref{sec:repeated_halving_algorithm}, and the entirety of our work is summarized in Section~\ref{sec:conclusion}.

\section{Related Work}\label{sec:related_work}

{\bf Leverage Score}
 We use  $\sigma_i(A)$ to  denote the leverage score of the $i$-th row as $\sigma_i(A)$, defined as $\sigma_i(A) = a_i^\top (A^\top A)^\dagger a_i$ given $A \in \R^{n \times d}$. The concept of leverage scores finds extensive applications in graph theory and linear algebra.

In graph theory, leverage scores play a crucial role in various applications such as maximum matching \cite{bln+20, lsz20}, max-flow problems \cite{ds08, m13, m16, ls20}, graph sparsification \cite{s11}, and the generation of random spanning trees \cite{s18}.

Within the field of optimization, leverage scores are heavily explored in linear programming \cite{ls14, blss20}, the approximation of the John Ellipsoid \cite{ccly19}, cutting-plane methods \cite{v89, lsw15}, and semi-definite programming \cite{jkl+20}. In matrix analysis, \cite{swz19} investigates tensor CURT decomposition, while \cite{bw14,swz17,swz19} focus on matrix CUR decomposition. Other works, such as \cite{ss11, DMIMW12}, concentrate on approximating leverage scores. Simultaneously, the generalized concept of Lewis weight is explored by \cite{blm89, cp15}.

{\bf Quantum Sketching Algorithms}
In quantum computing, sketching seems unnecessary because the block-encoding framework can exponentially reduce the dimensions. For example, linear regression can be solved in poly-logarithmic time in quantum without using sketching \cite{wan17,cgj18,cd21,sha23}. However, as discussed in Section~\ref{sec:intro}, the quantum linear algebra and block-encoding-based approaches have several intrinsic limitations. This motivates us to investigate other quantum algorithmic paradigms for solving classical optimization and machine learning problems. Before our work, there are only a few works on quantum sketching algorithms. \cite{aw22} proposed efficient quantum algorithms for graph sparsification and solving Laplacian systems with some polynomial speedups. \cite{p14,lz17,sha23} designed quantum algorithms for leverage score approximation and sampling. However, these algorithms use quantum linear algebra, and the complexities depend on some data-dependent parameters (e.g., the condition number of the matrix). Very recently, \cite{ag23} proposed a quantum algorithm for solving linear programming using the interior-point method. A key sub-routine is a quantum algorithm for matrix spectral approximation, which quantizes the classical algorithm \cite{clm+15}.  Our work generalizes this result to Kronecker spectral approximation. 

\section{Preliminary}\label{sec:preli}

\subsection{Notations}

We use $\Pr[]$ to denote probability. We use $\E[]$ to denote expectation.

For vectors $u, v \in \R^d$ we let $\langle u, v \rangle = \sum_{i \in [d]} u_i v_i$ denote the standard inner product. A positive definite matrix $H$ defines an inner product $\langle u,v \rangle_H :=\langle u,Hv\rangle$.

For $\epsilon>0$, the matrix $Q \in S^d$ is an $\epsilon$-spectral approximation of a $d$-by-$d$ PSD matrix~$H \succeq 0$ if 
\[
(1-\epsilon) Q \preceq H \preceq (1+\epsilon) Q, 
\]
where $A \preceq B$ is equivalent to $B-A \succeq 0$.
We denote such a spectral approximation by $Q \approx_{\epsilon} H$.
We will also use this notation for scalars, where $a \approx_\epsilon b$ denotes $(1-\epsilon) b \leq a \leq (1+\epsilon) b$, and for vectors (where the inequalities hold entrywise).

Let $A \in \R^{m_1 \times n_1}$ and $B \in \R^{m_2 \times n_2}$. We define the Kronecker product between matrices $A$ and $B$, denoted $A \otimes B \in \R^{m_1 m_2 \times n_1 n_2}$, as $(A \otimes B)_{(i_1 - 1) m_2 + i_2, (j_1-1)n_2+j_2}$ 
is equal to $A_{i_1,j_1} B_{i_2,j_2}$, where $i_1 \in [m_1], j_1 \in [n_1], i_2 \in [m_2], j_2 \in [n_2]$. Given $m_1 = m_2 = m$ and $n_1 = n_2 = n$, we also use $\oplus$ to denote $A \oplus B \in \R^{m \times n}$, where $(A \oplus B)_{i,j} = A_{i,j} + B_{i,j}$.

In the context of a matrix $A \in \mathbb{R}^{m \times n}$, we use the notation $\| \cdot \|$ to denote spectral norms. Specifically, $\| \cdot \|_1$ corresponds to the entrywise $\ell_1$-norm defined as $\| A \|_1 = \sum_{i=1}^m \sum_{j=1}^n |A_{i,j}| $. We use \(\| A \|_2\) denotes the Frobenius-norm, i.e., $\|A\|_F := (\sum_{i=1}^m \sum_{j=1}^n |A_{i,j}|^2 )^{1/2}$. We include $\| \cdot \|_0$ to denote the $\ell_0$-norm, which counts the number of non-zero elements in the matrix.

For a square and full-rank matrix $A$, we use $A^{-1}$ to denote its true inverse. For any matrix $B$, we use $B^{\dagger}$ to denote the pseudo-inverse.

Given matrix $A$ and $B$, we use $B \subseteq_{p} A$ to denote subsampling $B$ from $A$ by keeping any individual row with a probability $p$.

We use $|i\rangle\in \C^{2^m}$ to represent the $i$-th computational basis quantum state in the $m$-qubit Hilbert space. 

``with high probability'' means with probability at least $1-1/n^c$ for an arbitrarily high but fixed constant $c>0$, where $n$ typically denotes the size of the problem instance.

\subsection{Basic linear algebra facts}

Due to the standard property of tensor $\otimes$, it is easy to observe the following fact,
\begin{fact}\label{fac:otimes_dot}
Given matrix $A , B, C, D$, we have
\begin{align*}
    (A \otimes B) (C \otimes D) = (A C) \otimes (B D).
\end{align*}
\end{fact}
\begin{fact}
If the following conditions hold
\begin{itemize}
    \item Let $A \in \R^{n \times n}$ denote a psd matrix
    \item Let $B \in \R^{n \times n}$ denote a psd matrix
    \item $ (1-\epsilon) B \preceq A \preceq (1+ \epsilon) B$
    \item Let $\epsilon \in (0,0.1)$
\end{itemize}
Then, we have
\begin{itemize}
    \item Part 1. $(1+\epsilon)^{-1} A \preceq B \preceq (1-\epsilon)^{-1} A$
    \item Part 2. $(1-2\epsilon) A \preceq B \preceq (1+2 \epsilon) A$
\end{itemize}
\end{fact}

\begin{fact}\label{fac:eigenvalue_spectral}
Let matrix $A \in \R^{n \times d}$ and $B \in \R^{n \times d}$, and let $\lambda(\cdot)$ represent eigenvalues. The spectrum of $A \otimes B$ equals $\{\lambda(A)_i \lambda(B)_j ~|~ i, j \in [D]\}$.
\end{fact}
The Chernoff bound lemma is introduced as follows.
\begin{lemma}[Chernoff bound]\label{lem:chernoff_bound}
    Let $Y = \sum_{i=1}^n X_i$, with $X_i$ the random variable indicating whether the $i$-th row of $A$ is in $A_L$. Then $\mu := \E[Y] \in \Theta(d)$. The multiplicative Chernoff bound for random variables in $\{0,1\}$ states that $\Pr[|Y-\mu| \geq \delta \mu] \leq 2 e^{-\delta^2 \mu/3}$ for $0 \leq \delta \leq 1$, and so $Y \in \Theta(d)$ except with probability $e^{-\Omega(d)}$.
\end{lemma}

We state the well-known JL Lemma as follows:
\begin{lemma}[JL Lemma \cite{jl86}] \label{lem:JL}
For all integers $n, D \geq 0$ and precision $\epsilon>0$, there exists a distribution ${\cal D}_{n,D,\epsilon}$ over matrices in $\R^{O(\log(n)/\epsilon^2) \times D}$ such that the following holds: for any set of vectors $x_i \in \R^D$, $i \in [n]$, it holds that if $\Pi \sim {\cal D}_{n,D,\epsilon}$ then with high probability
\begin{align*}
\| \Pi x_i \|_2^2
= (1\pm\epsilon) \| x_i \|_2^2, \quad \forall i \in [n].
\end{align*}
Moreover, the matrix $\Pi \sim {\cal D}_{n,D,\epsilon}$ can be sampled in time $\wt{O}(D/\epsilon^2)$.
\end{lemma}

\subsection{Fast Quantum Sampling Tool}

We first review the previous tool for one-dimensional fast quantum sampling (Lemma~\ref{lem:q_sampling_1_d}). Then, we prove a generalized version for multi-dimensional sampling (Lemma~\ref{lem:q_sampling_k_d}), which we believe is of independent interest.
\begin{lemma}[Quantum sampling in 1D, Claim 3  in \cite{aw22}, Lemma~3.10 in \cite{ag23}]\label{lem:q_sampling_1_d}
Let $p\in [0,1]^n$. There is a quantum algorithm such that 
\begin{itemize}
    \item {\bf Part 1.} Outputs the sampled elements where each element is sampled independently with probability $p_i\in [0,1]$.
    \item {\bf Part 2.} Runs in $\wt{O}(\sqrt{n\|p\|_1})$ time.
\end{itemize} 
\end{lemma}

\begin{lemma}[Quantum sampling in $k$-dimension]\label{lem:q_sampling_k_d}
If the following conditions hold
\begin{itemize}
    \item Let $n,k\in \mathbb{N}$.
    \item Let matrix $p \in [0,1]^{n \times k}$ and vectors $p^{(1)},\dots,p^{(k)}\in [0,1]^n$.
\end{itemize}
Then, there exists a quantum algorithm such that
\begin{itemize}
    \item {\bf Part 1.} Outputs the indices of the sampled elements where each element indexed by $(i_1,\dots,i_k)\in [n]^k$ is sampled independently with probability $\prod_{j=1}^k p^{(j)}_{i_j}$.
    \item {\bf Part 2.} It runs in time
    \begin{align*}
    \wt{O} ( \sqrt{n}\cdot \sum_{j=1}^k\sqrt{\|p^{(j)}\|_1}\cdot \prod_{l=1}^{j-1}\|p^{(l)}\|_1 ).
    \end{align*}
\end{itemize}
\end{lemma}
\begin{proof}
We simulate the sampling tasks in a tree structure. Intuitively, we sample the first coordinate $i_1$ at the first level according to  $p^{(1)}$. In expectation, about $\|p^{(1)}\|_1$ elements will be selected. Then, we sample the second coordinate $i_2$ for each selected element according to  $p^{(2)}$. In total, the tree has $k$ levels, and each node at level-$i$ has $\|p^{(i)}\|_1$ children. 

We can apply the quantum algorithm in Lemma~\ref{lem:q_sampling_1_d} to decide the children's indices. Then, the cost of a node at level-$j$ is $\wt{O}(\sqrt{n\|p^{(j)}}\|_1)$ for $j\in [k]$. 
And there are $\prod_{l=1}^{j-1} \|p^{(l)}\|_1$ nodes at level-$j$. Thus, the total cost is 
\begin{align*}
    & ~ \sum_{j=1}^k \prod_{l=1}^{j-1} \|p^{(l)}\|_1 \cdot  \wt{O}(\sqrt{n\|p^{(j)}}\|_1) \\
    = & ~ \wt{O}\Big( \sqrt{n}\cdot \sum_{j=1}^k\sqrt{\|p^{(j)}\|_1}\cdot \prod_{l=1}^{j-1}\|p^{(l)}\|_1 \Big).
\end{align*}
The lemma is then proved.
\end{proof}

\subsection{Quantum input model} \label{sec:random-oracle}
The input model of our quantum algorithm is the row-query model for the matrix:
\begin{align*}
    {\cal O}_A|i\rangle |0\rangle = |i\rangle |a_i\rangle~~~\forall i\in [n],
\end{align*}
where $a_i$ is the $i$-th row of $A$. The time complexity of each query is $O(d)$ or $O(r)$ if $A$ is $r$-sparse.

Our subsequent quantum repeated halving algorithm (refer to Algorithm~\ref{alg:quantum-halving:otimes}) needs to query a long classical random string efficiently. Similar to \cite{ag23}, we use the following lemma to remedy this issue at the price of using a QRAM, which is a strong assumption. We leave it as an open question to remove the QRAM.  
\begin{lemma}[Quantum random oracles Claim 1 in~\cite{aw22}] \label{lem:random-oracle}
Consider any quantum algorithm with runtime $q$ that makes queries to a uniformly random string.
We can simulate this algorithm with a quantum algorithm with runtime $\wt{O}(q)$ without random string, using an additional QRAM of $\wt{O}(q)$ qubits.
\end{lemma}

\subsection{Leverage Score Distribution}
 
We define the leverage score, which is a well-known concept in numerical linear algebra.
\begin{definition}[Leverage Score, see Definition B.28 in \cite{swz19} as an example]\label{def:leverage_score}
Given a matrix $A \in \R^{n \times d}$, let $U \in \R^{n \times d}$ denote the orthonormal basis of $A$. We define $\sigma_i := \| U_{i,*} \|_2^2$ for each $i \in [n]$. We say $\sigma$ is the leverage score of $A$.

It is well known that, leverage score alternatively can be defined as follows
\begin{align*}
\sigma_i(A) := a_i^\top (A^\top A)^{\dagger} a_i.    
\end{align*}
where $a_i^\top$ is the $i$-th row of $A$ for all $i \in [n]$
\end{definition}

\begin{fact}\label{fac:leverage_score_sum}
It is well known that $\sum_{i=1}^n \sigma_i = d$.
\end{fact}

\begin{definition}[$D \sim \mathsf{LS}(A)$, see Definition B.29 in \cite{swz19} as an example]\label{def:ls_distribution}
 Let $c > 1$ denote some universal constant. 

 For each $i \in [n]$, we define $p_i := c \cdot \sigma_i / d$.  

 Let $q \in \R^n$ be the vector that $q_i \geq p_i$.

 Let $m$ denote the sparsity of diagonal matrix $D \in \R^{n \times n}$.
 
 We say a diagonal matrix $D$ is a sampling and rescaling matrix according to leverage score of $A$ if for each $i \in [n]$, $D_{i,i} = \frac{1}{\sqrt{m q_i} }$ with probability $q_i$ and $0$ otherwise. (Note that each $i$ is picked independently and with replacement) We use notion $D \sim \mathsf{LS}(A)$ to denote that.
\end{definition}

\subsection{Matrix Subsampling}
\begin{definition}[Matrix Subsampling]\label{def:matrix_subsample}
The simplest way of subsampling is just keeping any individual row with a probability $p$.
We use 
\begin{align*}
B \subseteq_p A
\end{align*}
to denote the matrix obtained by keeping every individual row of $A$ independently with probability $p$.
    
\end{definition}

\begin{definition}[Weignted Subsampling (Refined version of Definition~\ref{def:matrix_subsample})]\label{def:weighted_supsampling}
    We will also use a more refined notion of subsampling.
For a matrix $A \in \R^{n \times d}$ and an entrywise positive weight vector $w \in \R^n = (w_i)_{i \in [n]}$, we denote by
\begin{align*}
B \overset{w,\epsilon}{\leftarrow} A
\end{align*}
\end{definition}

The following lemma states that a weighted subsampling of a matrix according to the leverage-score distribution implies spectral approximation. 
\begin{lemma}[Consequence of the matrix Chernoff bound~\cite{t12} 
]\label{lem:matrix-chernoff}
If the following conditions hold
\begin{itemize}
    \item Consider matrix $A \in \R^{n \times d}$
    \item  Weight vector $w = (w_i)_{i \in [n]}$ satisfying $w_i \geq \sigma_i(A)$
\item $B \overset{w,\epsilon}{\leftarrow} A$ be defined in Definition~\ref{def:weighted_supsampling}
\item Let $p_i =  \min\{1,c w_i \log(d)/\epsilon^2\}$, for $i \in [n]$.
\end{itemize}
Then we have
\begin{itemize}
    \item $B^\top B \approx_\epsilon A^\top A$ 
    \item $B$ has at most $O(\sum_{i \in [n]} p_i)$ rows
    \item It hold with high probability.
\end{itemize}
\end{lemma}

Now note that we always have 
\begin{align*}
\sum_{i \in [n]} p_i \leq c\log(d) \|w\|_1/\epsilon^2
\end{align*}
and $\sum_{i=1}^n \sigma_i(A) =d$ by Fact~\ref{fac:leverage_score_sum}.
Hence, if we have good estimates $w_i \in \Theta(\sigma_i(A))$ of the leverage scores of $A$, then $\|w\|_1 \in O(d)$ and the resulting matrix $B$ has only $O(d \log(d)/\epsilon^2)$ rows.

\subsection{Quantum Spectral Approximation for Matrices}

The following theorem proved in \cite{ag23} gives a quantum algorithm for matrix spectral approximation. Our work is mainly inspired by this result. Thus, we present a thorough review of it in Appendix~\ref{sec:review_q_alg}. 

\begin{theorem}[Quantum repeated halving algorithm, Theorem 3.1 in \cite{ag23}]\label{thm:q-rep-halving}
Consider query access to matrix $A \in \R^{n \times d}$ with row sparsity $r$. For any $\epsilon \in (0,1)$, there is a quantum algorithm that, returns a diagonal matrix $D \in \R^{n \times n}$ such that
\begin{itemize}
\item Part 1. $\| D \|_0 = O(\epsilon^{-2} d \log d)$
\item Part 2. $  (1-\epsilon) A^\top A \preceq A^\top D^\top D A \preceq (1+\epsilon) A^\top A $ 
\item Part 3. $D \sim \mathsf{LS}(A)$ (see Definition~\ref{def:ls_distribution})
\item Part 4. It makes $\wt{O}(\sqrt{nd} / \epsilon )$ row queries to $A$
\item Part 5. It takes $\wt{O}( r \sqrt{nd} / \epsilon + d^{\omega} )$ time
\item The success probability $0.999$
\end{itemize}
\end{theorem}

\section{Generalization of Leverage Scores}\label{sec:gen_leverage_score}
In our algorithm (See Algorithm~\ref{alg:quantum-halving:otimes}), the computation of leverage plays a crucial role. However, executing this computation incurs a significant computational cost. To address this challenge, we consider a generalization of leverage scores and efficiency leverage  score computation based on Johnson-Lindenstrauss (JL) (See Section~\ref{sec:efficient_leverage_computation}). This approach provides a robust estimate of leverage scores, which proves sufficient for our algorithm's requirements.
\begin{definition}[Generalization of leverage scores in \cite{cw17}]\label{def:generalize_leverage_score}
Let $A \in \R^{n \times d}, B \in \R^{n_B \times d}$.
For $i\in [n]$, we define the \emph{generalized} leverage score of the $i$-th row of a matrix $A$ with respect to a matrix $B$ as follows
\begin{align*}
\sigma^B_i(A)
:=\begin{cases}
a_i^\top (B^\top B)^\dagger a_i &\text{if } a_i \perp \ker(B) \\
\infty &\text{otherwise}.
\end{cases}
\end{align*}
\end{definition}

In the subsequent section, we will illustrate how the upper bound of the leverage score for matrix $A$ can be established through the generalized definition. Building upon this observation and in conjunction with Lemma~\ref{lem:uniform}, we can derive a spectral approximation of the matrix using leverage score-based sampling.
\begin{claim}
    If the following conditions hold:
    \begin{itemize}
        \item Let $\sigma_i(\cdot)$ be defined in Definition~\ref{def:generalize_leverage_score}
        \item Let $p \in [0, 1]$ and $B \subseteq_p A$ for $p \in [0,1]$.
        \item Let $\subseteq_p$ be defined in Definition~\ref{def:matrix_subsample}. 
        \item Let $A \in \R^{n \times d}$
    \end{itemize}
    We have that
    \begin{itemize}
        \item {\bf Part 1.} $\sigma_i(A) = \sigma^A_i(A)$
        \item {\bf Part 2.} $\sigma^B_i(A) \geq \sigma_i(A)$.
    \end{itemize}
\end{claim}

Our primary focus lies in the Kronecker product, a mathematical operation involving two matrices. Additionally, we will elucidate the establishment of spectral approximation for a single matrix through the utilization of a sampling method based on leverage scores. This explanation aims to enhance readers' comprehension of our methodology.

It is important to note that, despite the conceptual similarity, this method cannot be directly applied to our algorithm. Instead, we introduce a novel sampling approach specifically tailored for the quantum method proposed in Lemma~\ref{lem:q_sampling_k_d}. This method is presented in Section~\ref{sub:section:uniform_sampling} as one of our contributions, acknowledging Kronecker as a key element.

\begin{lemma}[Theorem~4 in \cite{clm+15}] \label{lem:uniform}
    If the following conditions hold 
\begin{itemize}
    \item $A \in \R^{n \times d}$ and $B \subseteq_{1/2} A$.
    \item $w \in \R^n$ by $w_i \geq \sigma^B_i(A)$
    \item $p_i =  \min\{1,c w_i \log(d)/\epsilon^2\}$
    \item $\widetilde B \overset{w,\epsilon}{\leftarrow} A$
\end{itemize}
We have 
\begin{itemize}
    \item {\bf Part 1.} $(1 - \epsilon) A^\top A\leq \widetilde  B^\top \widetilde B  \leq (1 + \epsilon) A^\top A$ 
    \item {\bf Part 2.} $\widetilde B$ has $O(d \log(d)/\epsilon^2)$ rows
    \item {\bf Part 3.} It holds with probability $\sum_{i \in [n]} p_i \in O(d \log(d)/\epsilon^2)$
\end{itemize}
\end{lemma}

\section{Repeated halving Algorithm For $2$-D version}\label{sec:repeated_halving_algorithm}
\subsection{Weighted Sampling Algorithm} 
For the purpose of simplifying the description in the following section, we provide an overview of 2D weighted subsampling here. Let $p^1 \in \R^n$ and $p^2 \in \R^n$ be given. From a quantum perspective, we efficiently sample from two matrices, ensuring that the time complexity remains no higher than sampling from a single matrix.

Simultaneously, we illustrate that sampling from $A_1$ and $A_2$ to obtain $B^1$ and $B^2$, respectively, where $B = B^1 \otimes B^2$, is equivalent to directly sampling from $A = A^1 \otimes A^2$ (resulting in a time complexity of $n^2$). In other words, $B^\top B \approx_\epsilon A^\top A$.

\begin{definition}[Weighted Sampling Algorithm]
Let algorithm be defined in Algorithm~\ref{alg:subsampling:2D}, we will use the following method to describe the alogrithm,i.e
\begin{align*}
    B_1, B_2 \overset{w^1,w^2}{\leftarrow}\mathrm{Sample}(A_1,A_2)
\end{align*}
\end{definition}
\begin{algorithm}[!ht]
\caption{Weighted Subsampling} \label{alg:subsampling:2D}
 \begin{algorithmic}[1]
\Procedure{Sampling}{$A_1 \in \R^{n\times d}$, $A_2 \in \R^{n \times d}$, $w^1 \in (\R_{>0} \cup \infty)^n$, $w^2 \in (\R_{>0} \cup \infty)^n$, $\epsilon>0$}
\For{$i \in [n]$}
\State with probability $p^1_i := \min\{1,c w^1_i \log(d)/\epsilon^2\}$, add row $\frac{1}{\sqrt{p_i}} a_i^\top$ to $B^1$\;
\State with probability $p^2_i := \min\{1,c w^2_i \log(d)/\epsilon^2\}$, add row $\frac{1}{\sqrt{p_i}} a_i^\top$ to $B^2$\;
\EndFor
\State \Return $B^1$, $B^2$
\EndProcedure
\end{algorithmic}
\end{algorithm}

\subsection{Uniform Sampling
}\label{sub:section:uniform_sampling}

\begin{theorem}[A $2$-D version of Lemma~\ref{lem:uniform}, Tensor version of Theorem~4 in \cite{clm+15}] \label{thm:uniform:2D}
If the following conditions hold 
\begin{itemize}
    \item $A^1 \in \R^{n \times d}$ and $A^2 \in \R^{n \times d}$
    \item $A = A^1 \otimes A^2 \in \R^{n^2 \times d^2}$
    \item $B_1 \subseteq_{1/2} A^1$ and $B_2 \subseteq_{1/2} A^2$.
    \item $w^1, w^2 \in \R^n$ by $w^1_i = \sigma^{B^1}_i(A^1)$ and  $w^2_i = \sigma^{B^2}_i(A^2)$
    \item $p^1_i =  \min\{1,c w^1_i \log(d)/\epsilon^2\}$ 
    \item $p^2_i =  \min\{1,c w^2_i \log(d)/\epsilon^2\}$ 
    \item $\widetilde B^1 \overset{w,\epsilon}{\leftarrow} A^1$, $\widetilde B^2 \overset{w,\epsilon}{\leftarrow} A^2$ (which sample from $A^1$ and $A^2$ independently)
    \item $\wt{B} = \widetilde B^1 \otimes \widetilde B^2$.
\end{itemize}
We have 
\begin{itemize}
    \item {\bf Part 1.} $(1 - \epsilon) A^\top A \preceq \widetilde  B^\top \widetilde B  \preceq (1 + \epsilon) A^\top A$ 
    \item {\bf Part 2.} $\widetilde B$ has $O( \epsilon^{-4} d^2 \log^2(d) )$ rows 
    \item {\bf Part 3.} It holds with probability 1 that $\sum_{i \in [n]} p_i \in O(d \log(d)/\epsilon^2)$
\end{itemize}
\end{theorem}
\begin{proof}

By Lemma~\ref{lem:uniform}, we have $(\wt{B}^1)^\top \wt{B}^1\approx_\epsilon (A^1)^\top A^1$ and $(\wt{B}^2)^\top \wt{B}^2\approx_\epsilon (A^2)^\top A^2$. 

Note that 
\begin{align*}
    A^\top A 
    = & ~ (A^1\otimes A^2)^\top (A^1\otimes A^2) \\
    = & ~ ((A^1)^\top \otimes (A^2)^\top) (A^1\otimes A^2) \\
    = & ~ ((A^1)^\top A^1) \otimes ((A^2)^\top A^2),
\end{align*}
where the first step follows from the definition of $A$, and the second and third steps follow from the properties of the Kronecker product (see Fact~\ref{fac:otimes_dot}).

\begin{align*}
    \wt{B}^\top \wt{B} = ((\wt{B}^1)^\top \wt{B}^1) \otimes ((\wt{B}^2)^\top \wt{B}^2).
\end{align*}

We also note that the spectrum of $A\otimes B$ is $\{\lambda(A)_i\cdot \lambda(B)_j~|~i,j\in [D]\}$, where $\lambda(A)$ and $\lambda(B)$ are eigenvalues of $A$ and $B$, respectively (See Fact~\ref{fac:eigenvalue_spectral}). The spectral approximation of $\wt{B}^1$ guarantees that 
\begin{align*}
    \lambda((\wt{B}^1)^\top \wt{B}^1)_i\in (1\pm \epsilon)\cdot \lambda((A^1)^\top A^1)_i~~~\forall i\in [D].
\end{align*}
And for $\wt{B}^2$, it holds that
\begin{align*}
    \lambda((\wt{B}^2)^\top \wt{B}^2)_j\in (1\pm \epsilon)\cdot \lambda((A^2)^\top A^2)_j~~~\forall j\in [D].
\end{align*}
Thus, for any $(i,j)\in [D]\times [D]$,
\begin{align*}
    & ~ \lambda((\wt{B}^1)^\top \wt{B}^1)_i \cdot \lambda((\wt{B}^2)^\top \wt{B}^2)_j \\ 
    \in & ~ (1\pm \epsilon)^2 \cdot \lambda((A^1)^\top A^1)_i\cdot \lambda((A^2)^\top A^2)_j\\
    \in & ~ (1\pm O(\epsilon)) \cdot \lambda((A^1)^\top A^1)_i\cdot \lambda((A^2)^\top A^2)_j.
\end{align*}
It implies that
\begin{align*}
    ((\wt{B}^1)^\top \wt{B}^1) \otimes ((\wt{B}^2)^\top \wt{B}^2) \approx_{O(\epsilon)} ((A^1)^\top A^1) \otimes ((A^2)^\top A^2).
\end{align*}
That is, $\wt{B}^\top \wt{B} \approx_{O(\epsilon)} A^\top A.$
 
\end{proof}

\subsection{Correctness of Repeated halving algorithm}

\begin{lemma}[Repeated halving Algorithm for $2$-D version] \label{lem:repeated-halving_2d}
If the following conditions hold
\begin{itemize}
    \item Let repeated halving algorithm be defined in Algorithm~\ref{alg:repeated-halving}
    
    \item Let $A = A^1 \otimes A^2 \in \R^{n^2 \times d^2}$ where $A^1 \in \R^{n \times d}$ and $A^2 \in \R^{n \times d}$.
    \item Let $\wt{B} = \wt{B}^1 \otimes \wt{B}^2$
\end{itemize}
With high probability,  we have that
\begin{itemize}
    \item {\bf Part 1.} $B_\ell$ has $O(d\log(d))$ rows
    \item {\bf Part 2.} 
    $\wt{B}^\top \wt{B} \approx_\epsilon A^\top A$
    \item {\bf Part 3.} 
    the output $\wt{B}$ has $O(d \log(d)/\epsilon^2)$ rows 
\end{itemize}
\end{lemma}
\begin{proof}
{\bf Proof of Part 1 and Part 2.}
Now, we will consider $A^1$ first.
First we prove that for every $1 \leq \ell < L$, with high probability, $B_\ell$ has $O(d \log(d))$ rows and $B_\ell^\top B_\ell \approx_{1/2} A_\ell^\top A_\ell$.

We prove this by induction.

The base case $\ell = L$ and $B^1_L = A^1_L$ directly implies that
\begin{align*}
    (B^1_L)^\top B^1_L \approx_{1/2} (A^1_L)^\top A^1_L
\end{align*}
and by a multiplicative Chernoff bound (See Lemma~\ref{lem:chernoff_bound}), $A_L$ has $\Theta(d)$ rows with high probability.

For the inductive step $l-1, l-2, \cdots, 1$, given that $B_{\ell+1}^\top B_{\ell+1} \approx_{1/2} A_{\ell+1}^\top A_{\ell+1}$ and Definition~\ref{def:generalize_leverage_score}, we have that 
\begin{align*}
 (1 - 1/2) \sigma^{A_{\ell+1}}_i(A_\ell) \leq   \sigma^{B_{\ell+1}}_i(A_\ell) \leq (1 + 1/2) \sigma^{A_{\ell+1}}_i(A_\ell)
\end{align*}
Let $w_i$ be defined in Line~\ref{line:B_ell} in Algorithm~\ref{alg:repeated-halving}. and it follows that 
$\sigma_i^{A_{\ell+1}}(A_\ell) \leq  w_i \leq 6 \sigma_i^{A_{\ell+1}}(A_\ell)$.
Hence we can apply Lemma~\ref{lem:uniform} with $\epsilon = 1/2$, which implies that with high probability $B_\ell^\top B_\ell \approx_{1/2} A_\ell^\top A_\ell$ and $B_\ell$ has $O(d \log d)$ rows. The proof of ${\bf Part 1}$ is finished.

Now we have that $B_1 \approx_{1/2} A_1$ and $\sigma_i^{A^1_{1}}(A) \leq  w_i \leq 6 \sigma_i^{A^1_{1}}(A)$.

Similarly, we also have that $ \sigma_i^{A^2_{1}}(A) \leq  w_i \leq 6 \sigma_i^{A^2_{1}}(A)$.
 
Notice that $\wt{B}^1 \overset{w,\epsilon}{\leftarrow} A^1$ and $\wt{B}^2 \overset{w,\epsilon}{\leftarrow} A^2$ and then by Theorem~\ref{thm:uniform:2D}, we have that $(1 - \epsilon) (A^1 \otimes A^2)^\top (A^1 \otimes A^2)\leq (\wt{B}^1 \otimes \wt{B}^2)^{\top} (\wt{B}^1 \otimes \wt{B}^2) $.
 
and 
$(\wt{B}^1 \otimes \wt{B}^2)^{\top} (\wt{B}^1 \otimes \wt{B}^2) \leq (1+\epsilon) (A^1 \otimes A^2)^\top (A^1 \otimes A^2)$
 
In addition, we know $\wt{B}$ has $O(d^2 \log(d)/\epsilon^2)$ rows.

Now the proof of {\bf Part 2} and {\bf Part 3} are finished.
\end{proof}

\section{Quantum Halving Algorithm}
\subsection{Time Complexity: Single Step For Quantum halving algorithm}
\begin{lemma}[Single Step For Quantum halving algorithm] \label{lem:single-iteration:A_1_otimes_A_2}
If the following conditions hold
\begin{itemize}
    \item Let $A^1 \in \R^{n \times d}$ and $A^2 \in \R^{n \times d}$ with row sparsity $r$.
    \item Let $A = A^1 \otimes A^2 \in \R^{n^2 \times d^2}$.
    \item Let $\epsilon \in (0,1]$
    \item Let the Sampling process be described in Algorithm~\ref{alg:subsampling:2D}.
    \item Consider a random submatrix $A' \subseteq_{1/2} A$ and assume that we are given $B' \in \R^{O(d \log d) \times d}$, with row sparsity $r$, such that $B'^\top B' \approx_{1/2} A'^\top A'$.
\end{itemize}
there is a quantum algorithm such that 
\begin{itemize}
    \item {\bf Part 1.} returns a matrix $B^1 \overset{w,\epsilon}{\leftarrow} A^1$ for $w_i = \min\{1,\tilde w_i\}$ with $2\sigma^{B'}_i(A^1) \leq \tilde w_i \leq 4\sigma^{B'}_i(A^1)$.
    \item {\bf Part 2. } returns a matrix $B^2 \overset{w,\epsilon}{\leftarrow} A^2$ for $w_i = \min\{1,\tilde w_i\}$ with $2\sigma^{B'}_i(A^2) \leq \tilde w_i \leq 4\sigma^{B'}_i(A^2)$.
     \item {\bf Part 3. Time Complexity.} The algorithm takes $\wt{O}(\sqrt{nd}/\epsilon)$ row queries to $A^1$ and $A^2$ and $\wt{O}(r \sqrt{nd}/\epsilon + d^\omega)$ time.
    \item {\bf Part 4.} It holds with high probability.
\end{itemize}
\end{lemma}
\begin{proof}
    {\bf Proof of Part 1 and Part 2.}
For {\bf Part 1} and {\bf Part 2}, we will consider $A^1$ first. And $A^2$ will share the similar proof.

First we apply Lemma~\ref{lem:approx-LS} to $A^1$ and $B'$, with $D = O(d \log(d))$.
It states that, after a precomputation time of $\wt{O}(d^\omega)$, we can query approximate leverage scores $\tilde\sigma_i$ satisfying $\tilde\sigma_i \approx_{1/2} \sigma_i^{B'}(A^1)$ for $i \in [n]$, by querying one row of $A$ and using time $O(r_{A^1} \log(n))$ per leverage score.
Now set $w_i = \min\{1,\tilde w_i\}$ with $\tilde w_i = 2 \tilde\sigma_i$, which satisfies $2\sigma^{B'}_i(A^1) \leq \tilde w_i \leq 4\sigma^{B'}_i(A^1)$.

{\bf Proof of Part 3 and Part 4.}
We define the list $p = (p_i)_{i \in [n]}$ by setting 
\begin{align*}
p_i = \min\{1,c w_i \log(d)/\epsilon^2\}
\end{align*}(See Section~\ref{sec:rep-halving}), and so we can query $p_i$ at the same cost of querying~$\tilde\sigma_i$.

To sample $B^1 \overset{w,\epsilon}{\leftarrow} A^1$, we apply the algorithm from Lemma~\ref{lem:q_sampling_1_d} to the list $p$.
Lemma~\ref{lem:uniform} implies that $\sum_i p_i \in O(d \log(d)/\epsilon^2)$, and so the complexity amounts to $\wt{O}(\sqrt{n d}/\epsilon)$ queries to $p$.
$A^2$ follows the same proof above.

Combined with the precomputation time of $\wt{O}(d^\omega)$ and Lemma~\ref{lem:q_sampling_1_d}, this gives a total complexity of $\wt{O}((r_{A^1} + r_{A^2}) \sqrt{nd}/\epsilon + d^\omega)$ time and $\wt{O}(\sqrt{nd}/\epsilon)$ row queries to $A^1$ and $A^2$.
\end{proof}

\subsection{Main Result: Quantum Repeated halving algorithm}
\begin{algorithm}[!ht]
\caption{Quantum Kronecker repeated halving algorithm:} \label{alg:quantum-halving:otimes}
\begin{algorithmic}[1]
\Procedure{QuantumOtimes}{$A^1 \in \R^{n \times d}$, $A^2 \in \R^{n \times d}$, $\epsilon>0$}
 
\State implicitly construct the chain $A^1_L \subseteq_{1/2} \dots \subseteq_{1/2} A^1_1 \subseteq_{1/2} A^1$ for $L = O(\log(n/d))$\; \label{line:sample_A}
\State implicitly construct the chain $A^2_L \subseteq_{1/2} \dots \subseteq_{1/2} A^2_1 \subseteq_{1/2} A^2$ for $L = O(\log(n/d))$\; \label{line:sample_B}

\State use Grover search\footnotemark{} to explicitly learn $A^1_L, A^2_L$
\State $B_L \leftarrow A_L$\;
\For{$\ell = L-1,L-2,\dots,1$}
\State Construct $B^1_\ell \overset{w,1/2}{\leftarrow} A^1_\ell$ with $w_i^1 = \min\{1,\tilde w_i^1\}$ and $2\sigma^{B^1_{\ell+1}}_i(A^1_\ell) \leq \tilde w_i^1 \leq 4\sigma^{B^1_{\ell+1}}_i(A^1_\ell)$\;
\State Construct $B^2_\ell \overset{w,1/2}{\leftarrow} A^2_\ell$ with $w_i^2 = \min\{1,\tilde w_i^2\}$ and $2\sigma^{B^2_{\ell+1}}_i(A^2_\ell) \leq \tilde w_i^2 \leq 4\sigma^{B^2_{\ell+1}}_i(A^2_\ell)$\; \Comment{Use Lemma~\ref{lem:q_sampling_1_d} to run the following sampling steps in the same time}
\EndFor
\State  Construct $B^1 \overset{w,\epsilon}{\leftarrow} A^1$ with $w_i^1 = \min\{1,\tilde w_i^1\}$ and $2\sigma^{B^1_1}_i(A^1) \leq \tilde w_i^1 \leq 4\sigma^{B^1_1}_i(A^1)$\;
\State  Construct $B^2 \overset{w,\epsilon}{\leftarrow} A^2$ with $w_i^2 = \min\{1,\tilde w_i^2\}$ and $2\sigma^{B^2_1}_i(A^2) \leq \tilde w_i^2 \leq 4\sigma^{B^2_1}_i(A^2)$\; \Comment{Use Lemma~\ref{lem:q_sampling_1_d} to run the following sampling steps in the same time}
\State \Return $B^1 \otimes B^2$
\EndProcedure
\end{algorithmic}
\end{algorithm}

\begin{theorem}[Main Result, a formal version of Lemma~\ref{lem:otimes_quantum}]
If the following conditions hold
\begin{itemize}
    \item Let $A_1 \in \R^{n \times d}, A_2 \in \R^{n \times d}$.
    \item $A (A = A_1 \otimes A_2)\in \R^{n^2 \times d^2}$.
    \item $A$ with row sparsity $r$ and $\epsilon \in (0,1)$
\end{itemize}
there is a quantum algorithm (See Algorithm~\ref{alg:quantum-halving:otimes}) that such that
\begin{itemize}
    \item {\bf Part 1.} returns the sparse representation of a diagonal matrix $D \in \R^{n^2 \times n^2}$ (Each row in $DA$ is a scaling of one of rows in $A$. $D \sim \mathsf{LS}(A)$.)
    \item {\bf Part 2.} $\| D \|_0 = O(\epsilon^{-2} d^2 \log d)$
    \item {\bf Part 3.} $ (1-\epsilon) A^\top A \preceq A^\top D^\top D A \preceq (1+\epsilon) A^\top A $
    \item {\bf Part 4.} It makes $\wt{O}(\sqrt{nd} / \epsilon )$ row queries to $A$ and takes $\wt{O}( r \sqrt{nd} / \epsilon + d^{\omega} )$ time
    \item {\bf Part 5.} The success probability $0.999$
\end{itemize}
\end{theorem}

\begin{proof}
    {\bf Proof of Part 1.} It directly follows from Lemma~\ref{lem:repeated-halving_2d}.
    
    {\bf Proof of Part 2 and Part 3.}
    It directly follows from Lemma~\ref{lem:single-iteration:A_1_otimes_A_2}
\end{proof}

\section{Conclusion}\label{sec:conclusion}
As a widely utilized operator in linear algebra, the Kronecker product plays a pivotal role. Owing to its computationally intensive nature, researchers have delved into employing spectral approximation as an efficient calculation method. However, the conventional approach to performing such operations entails a time complexity squared compared to that of a single matrix without considering the Kronecker product. In response to this challenge, we propose a novel method that reduces the time complexity to $O_{d,\epsilon}(\sqrt{n})$ employing a quantum methodology. Our aspiration is that our work will attract more researchers to engage in the application of quantum ideas within the realm of classical algorithm theory.

\ifdefined\isarxiv

\else

\section*{Impact Statement}

This document introduces research aimed at pushing the boundaries of the Machine Learning field. Our work may have various societal implications, but we choose not to emphasize any specific consequences in this context.
\fi
\ifdefined\isarxiv
\else
\bibliography{ref}
\bibliographystyle{icml2024}

\fi

\newpage
\onecolumn
\appendix
\section*{Appendix}

{\bf Roadmap.}
In Section~\ref{sec:review_q_alg}, we present the classical algorithm for spectral approximation of a single matrix, employing the repeated halving algorithm. Additionally, we discuss its quantum counterpart. We also introduce the efficient computation of leverage scores, a key component repeatedly utilized in our algorithm.

\section{Quantum Algorithm for Matrix Spectral Approximation}\label{sec:review_q_alg}
For self-containedness, we review the quantum algorithm proposed by \cite{ag23} that achieves a quantum speedup for matrix spectral approximation. It mainly consists of the following three components:
\begin{enumerate}
    \item Classical repeated halving framework proposed in \cite{clm+15}.
    \item Classical efficient approximation of leverage scores via JL lemma.
    \item Quantum implementation of the repeated halving procedure.
\end{enumerate}
We introduce these components in the following three subsections.

\subsection{Repeated Halving Algorithm}\label{sec:rep-halving}
 The matrix size is reduced by $A' \subseteq_{1/2} A$ by a factor roughly two by uniform subsampling.  
In the subsequent algorithm (See Algorithm 2 in \cite{clm+15}), we obtain a sequence of $L \in O(\log(n/d))$ uniformly downsampled matrices, denoted as $A_1, \dots, A_L$. This sequence reaches a matrix $A_L$ with a mere $\wt{O}(d)$ rows. By Lemma~\ref{lem:uniform}, we can extract leverage scores $A_\ell$ (See Section~\ref{sec:gen_leverage_score}) to construct an approximate matrix $B_{\ell-1}$ corresponding to $A_{\ell-1}$. Replicating this procedure for $L$ iterations eventually produces an approximation $B_0$ for the original matrix $A_0 = A$.

\begin{algorithm}[!ht]
\caption{Repeated halving algorithm} \label{alg:repeated-halving}
\begin{algorithmic}[1]
\Procedure{RepeatedHalving}{$A \in \R^{n \times d}$, $\epsilon>0$}
\State let $A_L \subseteq_{1/2} \dots \subseteq_{1/2} A_1 \subseteq_{1/2} A$ for $L = \lceil \log_2(n/d) \rceil$
\State $B_L \leftarrow A_L$\;
\For{$\ell = L-1,L-2,\dots,1$}
\State let $B_\ell \overset{w,1/2}{\leftarrow} A_\ell$ with $2\sigma^{B_{\ell+1}}_i(A_\ell) \leq w_i \leq 4\sigma^{B_{\ell+1}}_i(A_\ell)$\; \label{line:B_ell}
\EndFor
\State let $\wt{B} \overset{w,\epsilon}{\leftarrow} A$ with  $2\sigma^{B_1}_i(A) \leq  w_i \leq 4\sigma^{B_1}_i(A)$\;
\State \Return $\wt{B}$ 
\EndProcedure
\end{algorithmic}
\end{algorithm}

By employing the concepts presented in \cite{clm+15}, we can derive the subsequent result.
\begin{lemma}[Repeated halving Algorithm] \label{lem:repeated-halving}
If the following conditions hold
\begin{itemize}
    \item Let repeated halving algorithm be defined in Algorithm~\ref{alg:repeated-halving}
    \item Let $A \in \R^{n \times d}$ be the input.
\end{itemize}
With high probability,  we have that
\begin{itemize}
    \item {\bf Part 1.} $B_\ell$ has $O(d\log(d))$ rows
    \item {\bf Part 2.} 
    $\wt{B}^\top \wt{B} \approx_\epsilon A^\top A$
    \item {\bf Part 3.} 
    the output $\wt{B}$ has $O(d \log(d)/\epsilon^2)$ rows 
\end{itemize}
\end{lemma}
\begin{proof}

First we prove that for every $1 \leq \ell < L$, with high probability, $B_\ell$ has $O(d \log(d))$ rows and $B_\ell^\top B_\ell \approx_{1/2} A_\ell^\top A_\ell$.

We prove this by induction.

The base case $\ell = L$ and $B_L = A_L$ directly implies that
\begin{align*}
    B_L^\top B_L \approx_{1/2} A_L^\top A_L
\end{align*}
and by a multiplicative Chernoff bound (See Lemma~\ref{lem:chernoff_bound}), $A_L$ has $\Theta(d)$ rows with high probability.

For the inductive step $l-1, l-2, \cdots, 1$, given that $B_{\ell+1}^\top B_{\ell+1} \approx_{1/2} A_{\ell+1}^\top A_{\ell+1}$ and Definition~\ref{def:generalize_leverage_score}, we have that 
\begin{align*}
 (1 - 1/2) \sigma^{A_{\ell+1}}_i(A_\ell) \leq   \sigma^{B_{\ell+1}}_i(A_\ell) \leq (1 + 1/2) \sigma^{A_{\ell+1}}_i(A_\ell)
\end{align*}
Let $w_i$ be defined in Line~\ref{line:B_ell} in Algorithm~\ref{alg:repeated-halving}. and it follows that 
$\sigma_i^{A_{\ell+1}}(A_\ell) \leq  w_i \leq 6 \sigma_i^{A_{\ell+1}}(A_\ell)$.
Hence we can apply Lemma~\ref{lem:uniform} with $\epsilon = 1/2$, which implies that with high probability $B_\ell^\top B_\ell \approx_{1/2} A_\ell^\top A_\ell$ and $B_\ell$ has $O(d \log d)$ rows. The proof of ${\bf Part 1}$ is finished.

Now we have that $B_1 \approx_{1/2} A_1$ and 
\begin{align*}
    \sigma_i^{A_{1}}(A) \leq  w_i \leq 6 \sigma_i^{A_{1}}(A)
\end{align*}
Given that $A_1 \subseteq_{1/2} A$ and by Lemma~\ref{lem:uniform} and $\wt{B} \overset{w,\epsilon}{\leftarrow} A$, it follows that 
\begin{align*}
    (1 - \epsilon) A^\top A\preceq \wt{B}^{\top}\wt{B} \preceq (1+\epsilon) A^\top A
\end{align*}
and 
$\wt{B}$ has $O(d \log(d)/\epsilon^2)$ rows.
 Now, the proof of {\bf Part 2 and 3} are finished.
\end{proof}

\subsection{Efficiency Leverage Score Computation}\label{sec:efficient_leverage_computation}

For any $\epsilon>0$, the computation of a multiplicative $(1\pm\epsilon)$-approximation of the generalized leverage scores $\sigma_i^B(A)$ is one of the main bottlenecks in the iterative halving algorithm is 

For the sake of completeness in our paper, we introduce an efficient algorithm for computing generalized leverage scores, which has been employed in various other works.
Here, $A \in \R^{n \times d}$ has $r_A$-sparse rows $a_i^\top$, and $B \in \R^{D \times d}$ has $r_B$-sparse rows, with $D \geq d$. Precisely calculating the generalized leverage scores, albeit naively, requires $O(r_A^2)$ time per score after some preprocessing.  
The Johnson-Lindenstrauss (JL) lemma allows us to obtain $\epsilon$-approximate leverage scores with a computational expense of $O(r_A \log(n)/\epsilon^2)$ per leverage score.

We start by rewriting the leverage scores as vector norms:
\begin{align*}
\sigma^{B}_i(A)
= & ~ a_i^\top (B^\top B)^{\dagger}  a_i \\
= & ~ a_i^\top (B^\top B)^{\dagger}  B^\top B (B^\top B)^{\dagger}  a_i \\
= & ~ \| B (B^\top B)^{\dagger}  a_i \|_2^2.
\end{align*}
Now, by Lemma~\ref{lem:JL}, we can sample a random matrix $\Pi \in \R^{O( \epsilon^{-2} \log(n) ) \times D}$ (independent of $A,B$)  such that with high probability (in $n$), 
\begin{align*}
\| \Pi B (B^\top B)^{\dagger}  a_i \|_2^2
= (1 \pm \epsilon) \| B (B^\top B)^{\dagger}  a_i \|_2^2,
\quad \forall i \in [n].
\end{align*}
Now, we can compute the matrix $C = \Pi B (B^\top B)^{\dagger}  \in \R^{O( \epsilon^{-2} \log(n) ) \times d}$ in additional time $O( \epsilon^{-2} d \log(n) \cdot (D+d))$ with the following steps:
\begin{itemize}
    \item Compute the $O(d\log(n)/\epsilon^2)$ entries of $\Pi B$, each of which is an inner product in dimension $D$
    \item  Compute $\Pi B (B^\top B)^{\dagger} $, this time with inner products in dimension $d$.
    \item Computate $\tilde\sigma_i = \| C a_i \|_2^2$ for the leverage scores, satisfying $\tilde\sigma_i = (1 \pm \epsilon) \sigma^{B}_i(A)$, at a cost per estimate of $1$ row-query to $A$ and time $O(r_A \log(n)/\epsilon^2)$.
\end{itemize}

Now, we have an efficiency  generalized leverage score computation algorithm as follows.
\begin{algorithm}[!ht]
\caption{Generalized leverage scores via Johnson-Lindenstrauss:} \label{alg:LS-via-JL}
\begin{algorithmic}[1]
\State matrices $A \in \R^{n \times d}$, $B \in \R^{D \times d}$, approximation factor $\epsilon>0$
\State \textbf{$-$ Preprocessing:}
\State sample random matrix $\Pi \sim {\cal D}_{n,D,\epsilon}$ and compute $C = \Pi B (B^\top B)^{\dagger} $\;
\State sample random matrix $\Pi' \sim {\cal D}_{n,d,\epsilon}$ and compute $C' = \Pi' (I - (B^\top B) (B^\top B)^{\dagger} )$\; 
\Procedure{LeverageScoreComputation}{$i \in [n]$}
\State Compute $C' a_i$; 
\If {$C' a_i \neq 0$} 
\State $\tilde\sigma_i \leftarrow \infty$\;
\Else
\State  $\tilde\sigma_i \leftarrow \| C a_i \|_2^2$\;
\EndIf
\State \Return $\tilde\sigma_i$
\EndProcedure
\end{algorithmic}
\end{algorithm}

Based on this algorithm we can prove the following lemma.

\begin{lemma}[Johnson-Lindenstrauss leverage scores] \label{lem:approx-LS}

If the following conditions hold
\begin{itemize}
    \item Let $A \in \R^{n \times  d}$.
    \item Let $i \in [n]$
    \item For each $i \in [n]$, let $\tilde\sigma_i$ be output of \textsc{LeverageScoreComputation}{($i\in [n]$)}.
    \item Let $r_B$-sparse matrix $B \in \R^{D \times d}$ with $n,D \gg d$.
    \item  Let preprocessing process be described in Algorithm~\ref{alg:LS-via-JL}.
    \item Let \textsc{LeverageScoreComputation} be described in Algorithm~\ref{alg:LS-via-JL}.
\end{itemize}
We have that
\begin{itemize}
    \item {\bf Part 1. Correctness} With high probability, $(1 - \epsilon)\sigma_i^B(A) \leq \tilde \sigma_i \leq (1 + \epsilon ) \sigma_i^B(A)$ 
    \item {\bf Part 2. Time Complexity of Prepocessing} The time complexity of prepocessing 
    \begin{align*}
    O(\min\{D d^{\omega-1}, D r_B^2 + d^\omega\} + \epsilon^{-2} d D\log(n) )\end{align*}.

    \item {\bf Part 3. Time Complexity} For $i \in [n]$, at a cost per estimate of one row query to $A$ and time $O(r_A \log(n)/\epsilon^2)$.
\end{itemize}
\end{lemma}
\begin{proof}

{\bf Proof of Part 1.}
First we show correctness of the algorithm.

By Lemma~\ref{lem:JL} and for all $i \in [n]$, it follows that
\begin{align*}
   (1 - \epsilon) \| B (B^\top B)^{\dagger}  a_i \|_2 \leq \| C a_i \|_2 \leq (1 +\epsilon) \| B (B^\top B)^{\dagger}  a_i \|_2
\end{align*}
and
\begin{align*}
 \| C' a_i \|_2 \geq & ~ (1 - \epsilon) \| (I - (B^\top B)(B^\top B)^{\dagger} ) a_i \|_2 \\
\| C' a_i \|_2 \leq & ~ (1 + \epsilon) \| (I - (B^\top B)(B^\top B)^{\dagger} ) a_i \|_2
\end{align*}
with high probability.

From $\| C' a_i \|_2$ we can now check whether $a_i \perp \ker(B)$ or not: indeed $I - (B^\top B) (B^\top B)^{\dagger} $ corresponds to the projector onto $\ker(B)$, and so $a_i \perp \ker(B)$ iff $\| C' a_i \|_2 = 0$.

Hence, if $\| C' a_i \|_2 = 0$ we can set $\tilde\sigma_i = \infty$ and otherwise we can set $\tilde\sigma_i = \| C a_i \|_2^2$, which will be correct with high probability.

{\bf Proof of Part 2}
Now we prove the complexity bounds.
First consider the precomputation phase.
As discussed earlier, given $B$ we can compute $(B^\top B)^{\dagger} $ and $(B^\top B)(B^\top B)^{\dagger} $ in time $O(\min\{D d^{\omega-1}, D r_B^2 + d^\omega\})$, and $C = \Pi B (B^\top B)^{\dagger} $ and $C' = \Pi' (I - (B^\top B) (B^\top B)^{\dagger} )$ in time $O(d D \log(n)/\epsilon^2)$.

{\bf Proof of Part 3.}
For the approximation of a single leverage score $\sigma^B_i(A)$, we query the entire row $a_i$ of $A$, and compute $C' a_i$ and $C a_i$ in time $O(r_A \log(n)/\epsilon^2)$. 
\end{proof}
 
\subsection{Quantum Implementation of Repeated Halving}

In this section, we introduce the quantum algorithm for obtaining a spectral approximation of a matrix via quantum repeated halving.

The following lemma shows the quantum speedup for a single iteration of the halving procedure.

\begin{lemma}[Quantum halving (one iteration)] \label{lemma:single-iteration}
If the following conditions hold
\begin{itemize}
    \item Let $A \in \R^{n \times d}$ with row sparsity $r$. 
    \item Let $\epsilon \in (0,1]$
    \item Consider a random submatrix $A' \subseteq_{1/2} A$ and assume that we are given $B' \in \R^{O(d \log d) \times d}$, with row sparsity $r$, such that $B'^\top B' \approx_{1/2} A'^\top A'$.
\end{itemize}
there is a quantum algorithm such that 
\begin{itemize}
    \item {\bf Part 1.} returns a matrix $B \overset{w,\epsilon}{\leftarrow} A$ for $w_i = \min\{1,\tilde w_i\}$ with $2\sigma^{B'}_i(A) \leq \tilde w_i \leq 4\sigma^{B'}_i(A)$.
     \item {\bf Part 2. Time Complexity.} The algorithm takes $\wt{O}(\sqrt{nd}/\epsilon)$ row queries to $A$ and $\wt{O}(r \sqrt{nd}/\epsilon + d^\omega)$ time.
    \item {\bf Part 3.} It holds with high probability.
\end{itemize}
\end{lemma}
\begin{proof}
{\bf Proof of Part 1.}
First we apply Lemma~\ref{lem:approx-LS} to $A$ and $B'$, with $D = O(d \log(d))$.
It states that, after a precomputation time of $\wt{O}(d^\omega)$, we can query approximate leverage scores $\tilde\sigma_i$ satisfying $\tilde\sigma_i \approx_{1/2} \sigma_i^{B'}(A)$ for $i \in [n]$, by querying one row of $A$ and using time $O(r_A \log(n))$ per leverage score.
Now set $w_i = \min\{1,\tilde w_i\}$ with $\tilde w_i = 2 \tilde\sigma_i$, which satisfies $2\sigma^{B'}_i(A) \leq \tilde w_i \leq 4\sigma^{B'}_i(A)$.

{\bf Proof of Part 2 and Part 3.}
We define the list $p = (p_i)_{i \in [n]}$ by setting 
\begin{align*}
p_i = \min\{1,c w_i \log(d)/\epsilon^2\}
\end{align*}(See Section~\ref{sec:rep-halving}), and so we can query $p_i$ at the same cost of querying~$\tilde\sigma_i$.

To sample $B \overset{w,\epsilon}{\leftarrow} A$, we apply the algorithm from Lemma~\ref{lem:q_sampling_1_d} to the list $p$.
Lemma~\ref{lem:uniform} implies that $\sum_i p_i \in O(d \log(d)/\epsilon^2)$, and so the complexity amounts to $\wt{O}(\sqrt{n d}/\epsilon)$ queries to $p$.

Combined with the precomputation time of $\wt{O}(d^\omega)$, this gives a total complexity of $\wt{O}(r_A \sqrt{nd}/\epsilon + d^\omega)$ time and $\wt{O}(\sqrt{nd}/\epsilon)$ row queries to $A$.
\end{proof}

By the classical repeated halving lemma (Lemma~\ref{lem:repeated-halving}), we can apply Lemma~\ref{lemma:single-iteration} for $O(\log(n))$ times. Then, with high probability, the returned matrix $B$ will have $O(d \log(d)/\epsilon^2)$ rows and satisfy $B \approx_\epsilon A$.

\begin{algorithm}[!ht]
\caption{Quantum repeated halving algorithm} \label{alg:quantum-halving}
\begin{algorithmic}[1]
\Procedure{QuantumRepeatedHalving}{$A \in \R^{n \times d}$, $\epsilon>0$} 
\State implicitly construct the chain $A_L \subseteq_{1/2} \dots \subseteq_{1/2} A_1 \subseteq_{1/2} A$ for $L = O(\log(n/d))$\; 
\State use Grover search\footnotemark{} to explicitly learn $A_L$
\State $B_L \leftarrow A_L$\;
\For{$\ell = L-1,L-2,\dots,1$}
\State use Lemma~\ref{lemma:single-iteration} on $A_{\ell+1} \subseteq_{1/2} A_\ell$ and $B_{\ell+1}$ to explicitly construct $B_\ell \overset{w,1/2}{\leftarrow} A_\ell$ with $w_i = \min\{1,\tilde w_i\}$ and $2\sigma^{B_{\ell+1}}_i(A_\ell) \leq \tilde w_i \leq 4\sigma^{B_{\ell+1}}_i(A_\ell)$\;
\EndFor
\State use Lemma~\ref{lemma:single-iteration} on $A_1 \subseteq_{1/2} A$ and $B_1$ to explicitly construct $B \overset{w,\epsilon}{\leftarrow} A$ with $w_i = \min\{1,\tilde w_i\}$ and $2\sigma^{B_1}_i(A) \leq \tilde w_i \leq 4\sigma^{B_1}_i(A)$\;
\State \Return $B$
\EndProcedure
\end{algorithmic}
\end{algorithm}

\footnotetext{More specifically, apply Lemma~\ref{lem:q_sampling_1_d} with $p_i = \Pi_{\ell \leq L} X^{(\ell)}_i$, where the $X^{(\ell)}_i$'s were defined in Section~\ref{sec:random-oracle}.}

\ifdefined\isarxiv
\bibliographystyle{alpha}
\bibliography{ref}
\else 

\fi




\end{document}